\documentclass{webofc}
\usepackage[varg]{txfonts}   
\usepackage{indentfirst}
\begin{document}
\title{Off-of-equilibrium effects on Kurtosis Along Strangeness-Neutral Trajectories}
%
%

\author{\firstname{Travis} \lastname{Dore}\inst{1}\fnsep \and
        \firstname{Jamie} \lastname{Karthein}\inst{2}\fnsep \and
        \firstname{Debora} \lastname{Mroczek}\inst{1}\fnsep \and
        \firstname{Paolo} \lastname{Parotto}\inst{3}\fnsep \and
        \firstname{Jacquelyn} \lastname{Noronha-Hostler}\inst{1}\fnsep \and
        \firstname{Claudia} \lastname{Ratti}\inst{2}\fnsep
}

\institute{Department of Physics, University of Illinois at Urbana-Champaign, Urbana, IL 61801, USA
\and
Department of Physics, University of Houston, Houston, TX 77204, USA
\and
Department of Physics, University of Wuppertal, Wuppertal D-42119, Germany
          }

\abstract{%
  The Beam Energy Scan program at  the Relativistic Heavy Ion Collider (RHIC)
is searching for the QCD critical point. The main signal for the
critical point is the kurtosis of the distribution of proton yields obtained on an event-by-event basis where one expects a peak at the critical point. However, its exact behavior is
still an open question due to out-of-equilibrium effects and uncertainty in the equation of state. Here we
use a simplistic hydrodynamic model that enforces strangeness-neutrality, selecting 
trajectories that pass close to the critical point. We vary the initial conditions to estimate the effect of out-of-equilibrium hydrodynamics on the kurtosis signal.
}
\maketitle
\vspace{-1cm}
\section{Introduction}
\label{intro}
As the search for the QCD critical point \cite{Aoki:2006br,Borsanyi:2010bp,Endrodi:2011gv,Bellwied:2015rza} continues, it becomes increasingly important to have a thorough understanding of its signals, including what factors influence them and how strong those influences are. One such signal is non-monotonic behavior of the kurtosis of the equilibrium net baryon distribution \cite{Mroczek:2020rpm,Stephanov:2011pb}, specifically, a large peak in the vicinity of the critical region signifying large fluctuations of net baryon number. Since the QCD equation of state (EoS) at finite baryon densities is not known from first-principles due to the Fermi sign problem, one must rely on phenomenological methods and models.
\par
\begin{figure}[h]
    \centering
    \begin{tabular}{cc}
       \includegraphics[clip=true,width=0.45\linewidth]{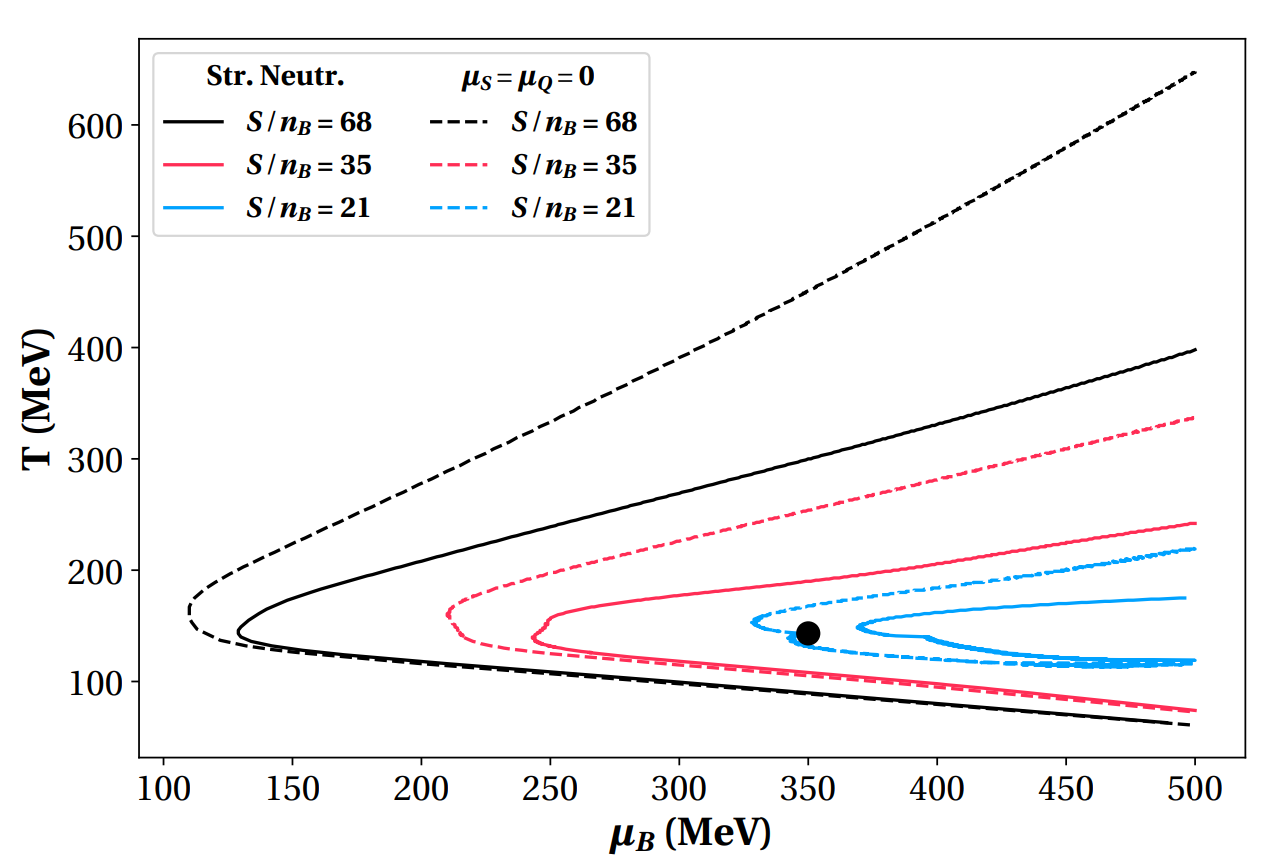}  &   \includegraphics[clip=true,width=0.5\linewidth]{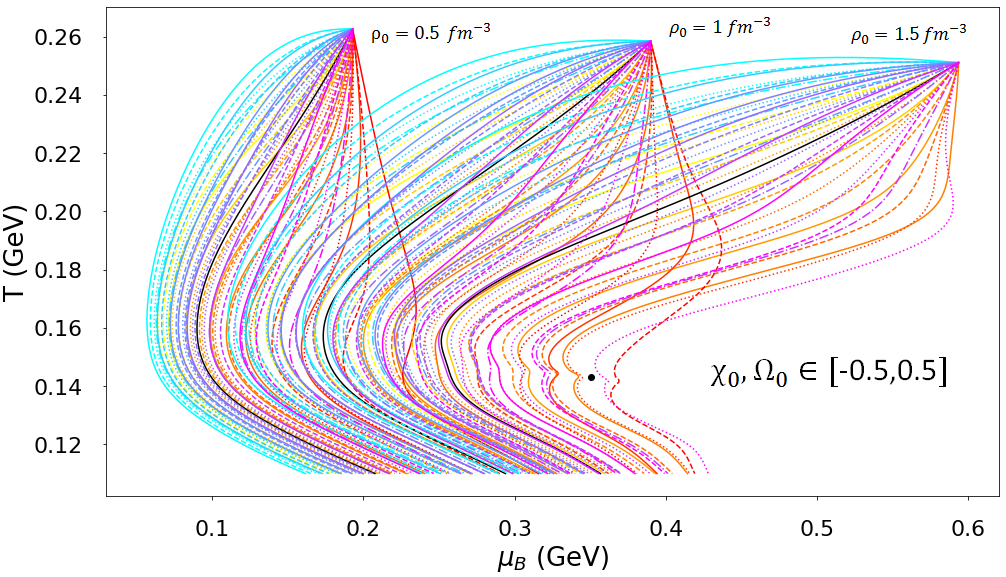}\\
    \end{tabular}
    \caption{Left: A comparison of isentropes from an EoS realization which enforces $\mu_B=\mu_S=0$ vs. one that enforces heavy-ion trajectories. Right: Demonstration of the ability of the system to pass through a large portion of the phase diagram depending on what its initial conditions are for its out-of-equilibrium components.}
    \label{fig:isenComp}
\end{figure}
One such implementation is done in Ref.\ \cite{Karthein:2021nxe} in which a critical region with the expected 3D Ising universality class is smoothly matched to the Lattice QCD EoS at zero net baryon density and uses a Taylor expansion for the non-critical contribution to the pressure at finite net baryon densities. The size, shape, and strength of the critical region can be selected through the use of multiple numerical parameters, each of which produce a realization of a QCD EoS with a critical point that can be constrained against experimental results and theoretical consistency checks (e.g. $0 \leq c_s^2 \leq 1 $). The aforementioned work is an extension of a previous one \cite{Parotto:2018pwx}, which now has the ability to capture `heavy-ion' trajectories through the phase diagram by enforcing both net-strangeness neutrality as well as finite electric charge, instead of $\mu_B=\mu_S=0$. The difference in the isentropes produced (which are trajectories in the phase diagram of constant entropy to baryon number ratio, $s/\rho_B$) by the two EoS's can be seen in Fig.\ \ref{fig:isenComp}.
\par
The EoS developed in that previous work is used here, with the parameterization being the same as that of the paper \cite{Karthein:2021nxe}. Here, we use this EoS in conjunction with a simplified hydrodynamic model to begin exploring out of equilibrium effects on the kurtosis at freeze-out. The value of the kurtosis at freeze-out is intimately related to the fluctuations in baryon number from event to event when the system freezes out \cite{Stephanov:1998dy,Stephanov:1999zu,Stephanov:2008qz,Stephanov:2011pb}, thus directly tying it to an observable. The work here does not use a realistic hydrodynamic model and thus cannot make quantitative predictions. Instead, these results are qualitative idea of the kinds of effects one might expect to encounter in realistic model-to-data comparisons.

\section{Out-of-Equilibrium Effects}
It has been shown in previous works \cite{Dore:2020jye,Du:2021zqz} that out of equilibrium effects can have an impact on the $(T,\mu_B)$ trajectories explored by the system throughout its evolution. This can have a direct influence on searches for the critical point by possibly pushing the system towards or away from the critical point on an event-by-event basis, and even affecting the path the system takes through the critical region. In Fig.\ \ref{fig:isenComp}, we demonstrate the ability of a system to pass through a large portion of the phase diagram, using the same energy density, three different baryon densities, and many different initial values for the shear-stress tensor and the bulk pressure. 

It is also expected that in the vicinity of the critical region the bulk viscosity, $\zeta/s$, should scale with the correlation length to the third power, i.e. $\zeta/s \sim (\xi/\xi_0)^3$ where $\xi_0$ is a scaling constant picked to scale the correlation length appropriately in the critical region \cite{Monnai:2016kud}. In order to incorporate as many critical effects as possible, we implement a bulk viscosity that has this critical scaling, the exact form of which, can be seen in the next section.
\section{Hydrodynamic Model}
\label{sec-1}
As a starting point for future work, we use a simplified Bjorken expanding model \cite{Bjorken:1982qr} to obtain qualitative results and a basic understanding of the sought after effects. Here we use the DNMR equations of motion detailed in \cite{Dore:2020jye}. These equations of motion require initial conditions for  $\pi_\eta^\eta$ - the rapidity eigenvalue of the shear stress tensor, $\Pi$ - the bulk pressure, the energy density $\epsilon$, and $\rho$ - the baryon density.  Second order transport coefficients are the same as in \cite{Denicol:2018wdp,Bazow:2016yra}.
We use a temperature dependent shear viscosity ($\eta T/w$ where $w=e+p$ is the enthalpy), given by an excluded volume calculation below the transition temperature \cite{NoronhaHostler:2012ug,McLaughlin:2021dph} (using the PDG16+ \cite{Alba:2017mqu}) that is matched onto a QCD motivated parameterization in the deconfined phase \cite{Christiansen:2014ypa,Dubla:2018czx}. The form of the non-critical and critical bulk viscosities are
 \begin{align}\label{eqn:zetanorm}
     \frac{\zeta T}{w} &=36\times \frac{1/3 - c_s^2}{8\pi}\\
     \left(\frac{\zeta T}{w}\right)_{CS} &=\frac{\zeta T}{w}
     \left[1 +\left(\frac{\xi}{\xi_0}\right)^3\right]
 \end{align}
 
 \section{Effects On Probed Kurtosis}
\begin{figure}
    \centering
    \includegraphics[width=5cm,clip]{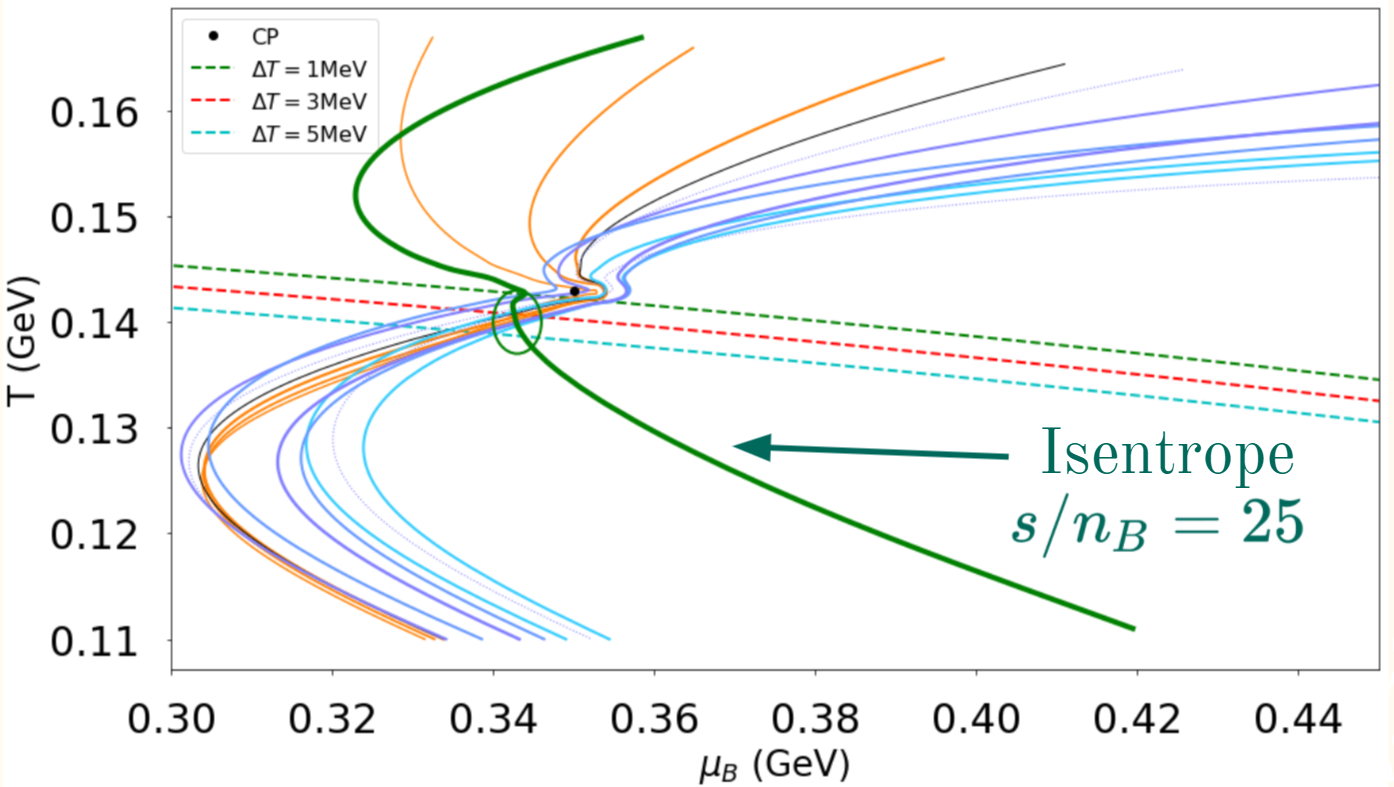}
    \caption{Trajectories of different calculations that pass through the critical region.}
    \label{fig:traj}
\end{figure}

\begin{figure} 
    \centering
    \begin{tabular}{ccc}
       \includegraphics[clip=true,width=0.33\linewidth]{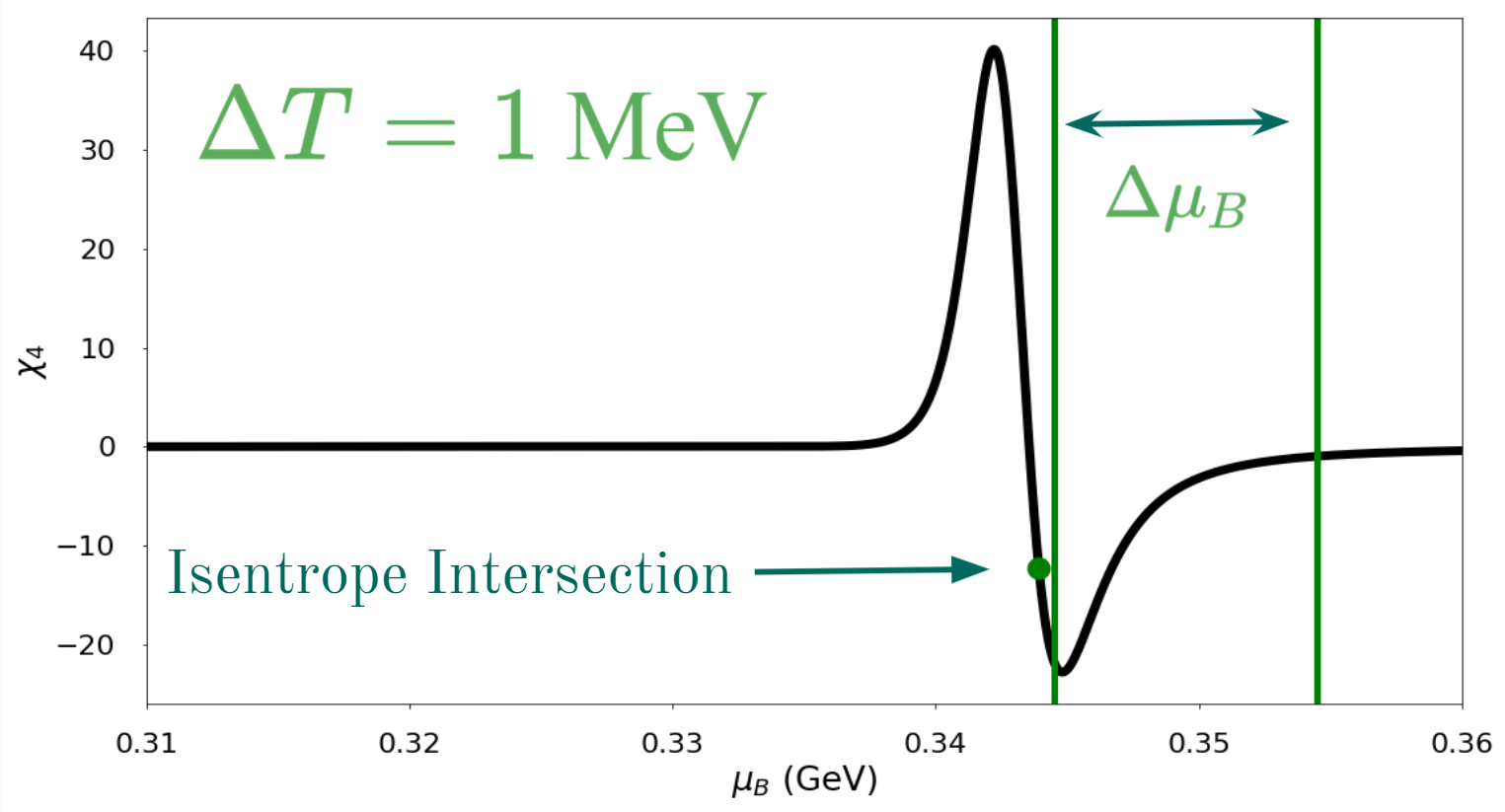}  &   \includegraphics[clip=true,width=0.33\linewidth]{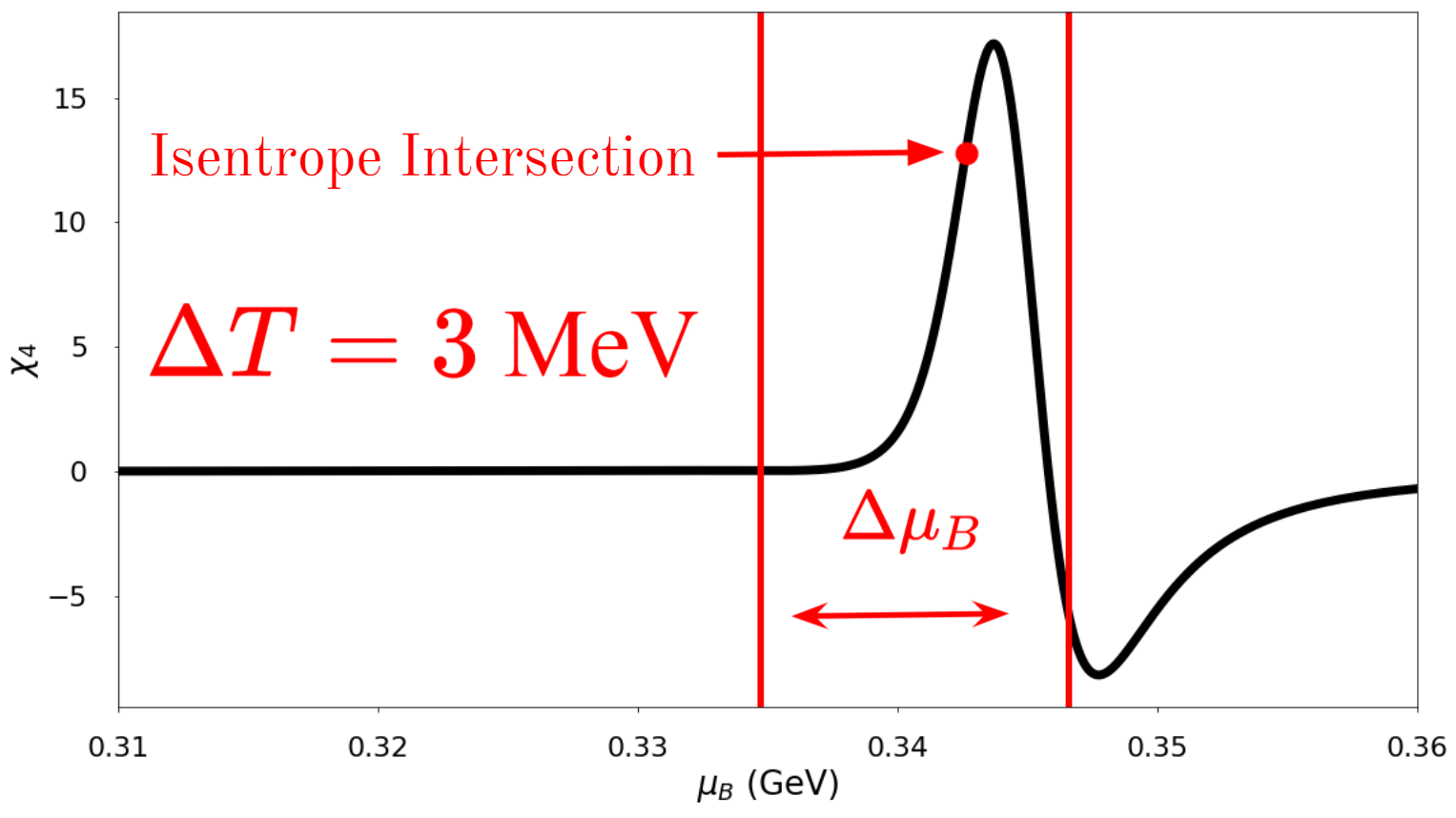} &   \includegraphics[clip=true,width=0.33\linewidth]{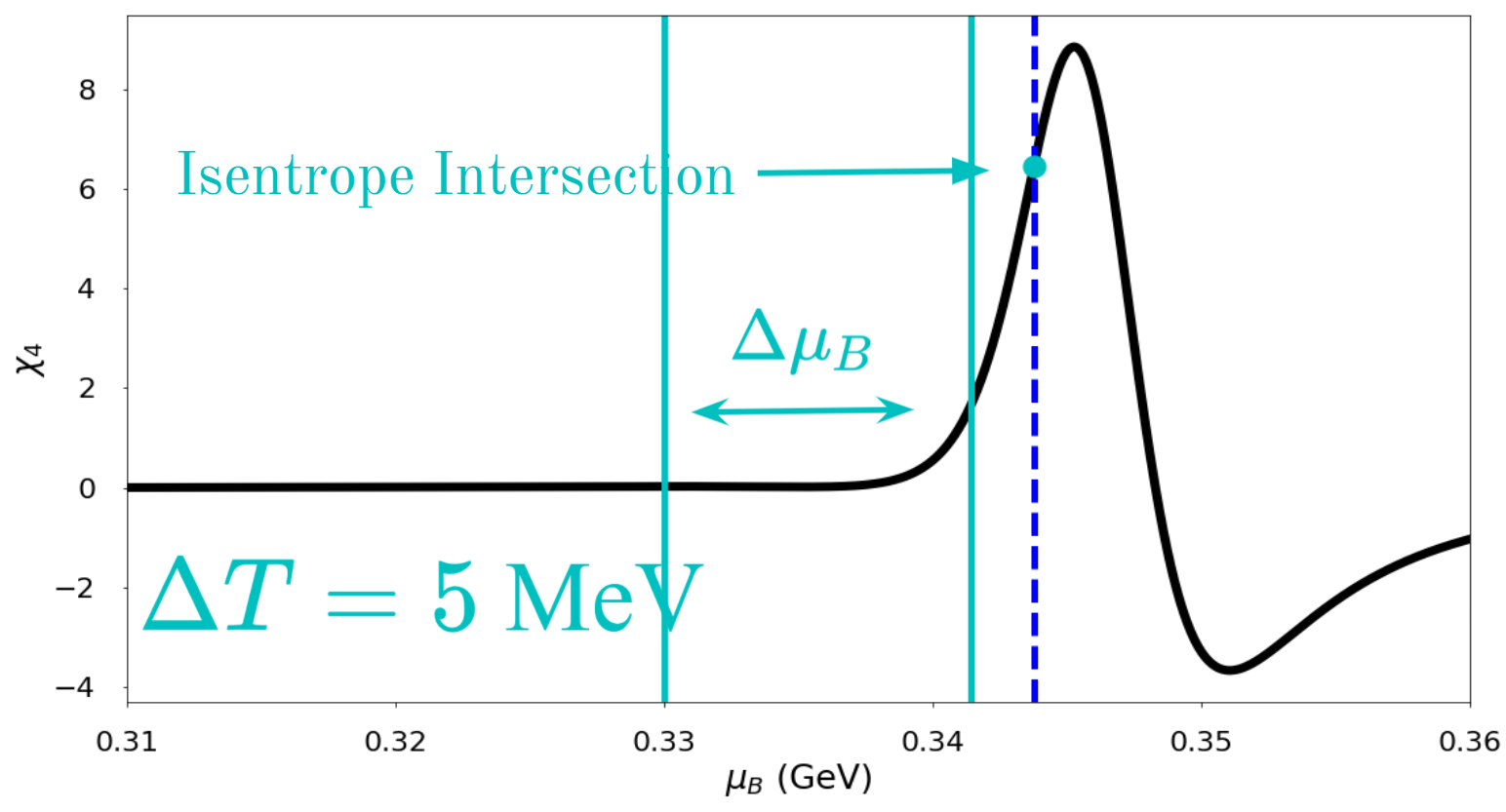}\\
    \end{tabular}
    \caption{The fourth moment of the equilibrium net baryon distribution at fixed values of temperature. The vertical lines are bands around possible values of the kurtosis seen including out of equilibrium effects. The dots are where the isentropes intersect the parabolas of constant temperature.}\label{fig:chi4}
 \end{figure}

In Fig.\ \ref{fig:traj}, one can see a variety of different hydrodynamic $(T,\mu_B)$ trajectories that the system can pass through and still go through the critical region. Here, we study which value of the kurtosis is actually probed by these runs, and check this at three different values of temperature after the critical point, specifically, 1, 3, and 5 MeV below $T_c$. In Fig.\ \ref{fig:chi4} we show which different values of $\chi_4$ are probed in our simulations. In this work, $\chi_4$ is used as a substitute for the kurtosis since the two are closely related.
 \par
 It can be seen in Fig.\ \ref{fig:chi4} that out of equilibrium effects lead to a spread in $\chi_4$ probed on an event-by-event basis. The vertical lines for each figure represent the bounds given by out-of-equilibrium effects in our work, the dots show where the isentrope probes $\chi_4$ at a given $(T,\mu_B)$. Interestingly, in the bottom right panel, the isentrope probes a region outside the bounds probed by out-of-equilibrium effects.

\section{Conclusion}
In this work we have explored some consequences that out-of-equilibrium effects could have on the search for the critical pointby looking at the fourth moment of the equilibrium net-baryon distribution. We found that, for a given EoS, the value of kurtosis you probe at freeze-out is heavily dependent on the out-of-equilibrium initial conditions of the system. This can lead to different values of the kurtosis being probed on an event-by-event basis.

This material is based upon work supported by the the US-DOE Nuclear Science Grant No. DE-SC0019175; the National Science Foundation under grants
no. PHY-1654219, PHY-1560077, and DGE – 1746047; and the DFG grant SFB/TR55. J.N.H. and T.D. are supported by the US-DOE Nuclear Science Grant No. DE-SC0020633. DM is partially supported by the National Science Foundation Graduate Research Fellowship Program under Grant No. DGE – 1746047

%
%
%
\bibliography{all}

\end{document}